\documentclass[12pt]{article}
\pdfoutput=1
\usepackage[numbers,sort&compress]{natbib}
\usepackage{graphicx}
\usepackage{amsmath,graphicx,amssymb,subfigure}
\usepackage[usenames,dvipsnames]{color}
\usepackage{amssymb}
\usepackage{amsmath}
\usepackage{isomath}
\usepackage{amsthm}
\usepackage{enumerate}
\usepackage{stmaryrd}
\usepackage{amsfonts}
\usepackage[all]{xy}
\usepackage[section]{placeins}
\usepackage[colorlinks=true,linktocpage=true,linkcolor=blue,citecolor=blue]{hyperref}
\newcommand{\mt}[1]{\textrm{\tiny #1}}
\newcommand{\be}{\begin{equation}}
\newcommand{\ee}{\end{equation}}
\newcommand{\bea}{\begin{eqnarray}}
\newcommand{\eea}{\end{eqnarray}}
\newcommand{\rh}{r_\mt{H}}
\newcommand{\n}{\nonumber}

\begin{document}

\bibliographystyle{hieeetr}

\pagestyle{plain}
\setcounter{page}{1}

\begin{titlepage}

\begin{center}

\vskip 30mm

{\Large\bf Charged  BTZ-like black hole solutions and the diffusivity-butterfly velocity relation}

\vskip 0.8cm

{\bf Xian-Hui Ge}$^{1}$,~ {\bf Sang-Jin Sin}$^{2}$,\\~ {\bf Yu Tian}$^{3}$,~ {\bf Shao-Feng Wu}$^{1}$,~{\bf Shang-Yu Wu}$^{4}$ \\
$^1${\it \small Department of Physics, Shanghai University, Shanghai 200444,  China}\\
$^2${\it\small Department of Physics, Hanyang University, Seoul  133-791, Korea}\\
$^3${\it\small School of Physics, University of Chinese Academy of Sciences, Beijing, 100049,  China}\\
$^4${\it  Department of Electrophysics, National Chiao Tung University, Hsinchu 300, Taiwan, ROC}\\
%{\sf{gexh@shu.edu.cn}}, {\sf{ytian@ucas.ac.cn}}, {\sf{sfwu@shu.edu.cn}}
\medskip

\vspace{5mm}
\vspace{5mm}

\begin{abstract}
 We show that there exists a class of charged BTZ-like black hole solutions in Lifshitz spacetime  with a hyperscaling violating factor.  The charged BTZ is characterized by a  charge-dependent logarithmic term in the metric function. As concrete examples, we give five such charged BTZ-like black hole solutions and the standard charged BTZ metric can be regarded as a special instance of them.  In order to check the recent proposed universal relations between diffusivity and the butterfly velocity, we first compute the diffusion constants of the  standard charged BTZ black holes and then extend our calculation to arbitrary dimension $d$, exponents  $z$ and  $\theta$.
 Remarkably, the case $d=\theta$ and $z=2$ is a very special in that the charge diffusion $D_c$ is a constant and the energy diffusion $D_e$ might be ill-defined, but $v^2_B\tau$ diverges.  We also compute the diffusion constants for the case that the DC conductivity is finite but in the absence of momentum relaxation.
\end{abstract}
\end{center}
\noindent
\end{titlepage}

\section{Introduction}

Black hole solutions are extremely useful in studying the holographic properties of strongly coupled quantum matters. For example, the Reissner-Nordstrom-Anti de Sitter black holes are of particular usage in the study of the electronic structure of the dual quantum systems \cite{Hartnoll09}. The holographic methods also yield disordered quantum systems with no quasiparticle excitations \cite{withers,blake}. However, the precise bulk dual of the recent proposed model, the Sachdev-Ye-Kitaev (SYK) model, is still elusive \cite{kiteav151,kiteav152}. The SYK models are theories of Majorana fermions in zero spatial dimensions with $q-$body infinite-range random interactions in Fock space \cite{pol16, mal16,jensen16}.

As one of the simplest strongly interacting system with a gravity dual, the SYK model has many interesting features including thermodynamic properties, correlation functions and the absence of quasiparticle in a non-trivial solvable limit in the presence of disorder at low temperature \cite{Yoon,Jevicki,Davison,Danshita,Gross,Gu,Berkooz,Gaiotto,Sachdev,Lamata,Hartnoll,Nishinaka,Turiaci,Jian,Chew,witten,Gurau,Peng,Klebanov,Ferrari,Itoyama,Peng2,Ludwig,Garca-Garca,Cotler,Nowak,Krishnan,Garca-Garca2,cai,Li,Xian}. All these properties suggest a gravity-dual interpretation of the model in the low-temperature strong-coupling limit and it is believed that the SYK models are connected holographically to black holes with $AdS_2$ horizons with non-vanishing entropy in the $T\rightarrow 0$ limit. It has been conjectured  that the bulk gravity dual of the SYK model is the two dimensional Jackiw-Teitelboim model of dilaton-gravity with a negative cosmological constant, while there are also some hints that it is actually Liouville theory. In \cite{1704.07208}, the authors show that the spectrum of the SYK model can be interpreted as that of a three-dimensional scalar coupled to gravity.  They further conjectured that the bulk dual of the  SYK model  is indeed a three dimensional theory.

 The main purpose of this paper is to study a class of charged Banados-Teitelboim-Zanelli-like ($\rm BTZ$-like) black hole solutions in general $d+2$-dimensional spacetime with a momentum dissipating source. The BTZ black hole sultions are asymptotically $AdS_3$ and can be dimensionally reduced to solutions of various two dimensional Jackiw-Teitelboim  theories. In \cite{2dcft}, a bulk theory with the BTZ black hole solution are utilized to study the late time behavior of the analytically continued partition function $Z(\beta+it)Z(\beta-it)$ in holographic $2d$ CFTS.  We will also compare the transport coefficients of such $\rm BTZ$-like solutions with those of the  SYK models.  Recently, aimed to build a connection between transport at strong coupling and quantum chaos, it has been conjectured that there is a universal relation between diffusion constant and the butterfly velocity $D\sim v^2_B\tau$ with a fundamental quantum thermal timescale $\tau\sim \hbar/k_B T$ \cite{shenker,blake15,hardy,stanford,sf1,sf2,niu}. In the holographic setup, the butterfly velocity in an isotropic system is defined as \cite{feng} $v^2_B=\frac{4\pi T g_{rr}}{d g'_{xx}(\rh)}\sqrt{g'_{tt}/g'_{rr}}$, where the prime indicates a radial derivative and $d$ is the dimension of spatial coordinates.  This relation is somehow inspired by the shear viscosity bound proposed by Kovtun, Son and Starinets \cite{kss}. But  recent developments have shown that the shear viscosity bound can be strongly violated in anisotropic systems and momentum dissipated systems \cite{cgs14,rebhan,cgs142,wge,gesc,GLNS2,ling16,fang16} (see also \cite{weijia16,baggioli17,weijia,lucas, fada} for discussions on the diffusion bounds ). In incoherent metals without a Drude peak, transports are dominated by diffusive physics in terms of charge and energy instead of momentum diffusion. One naturally guesses that the charge diffusion constant $D_c$ and energy diffusion constant $D_e$ could play a crucial role.   In the diffusion-butterfly effect scenario, the electron-phonon interactions of strongly correlated materials behave as a composite, strongly correlated soup with an effective velocity $v_B$ \cite{hardy}. A natural candidate for such a velocity is provided by the butterfly effect\cite{blake,hardy}.   We will study this relation for BTZ-like black holes. Interestingly, we also check the diffusivity-butterfly velocity relation in an extremal case: In a $d+2$-dimensional Lifshitz spacetime with a hyperscaling violating factor, the butterfly velocity behaves as $v^2_B=\frac{2\pi T \rh^{z-2}}{(d-\theta)}$. As $d\rightarrow \theta$, the butterfly velocity becomes divergent. Now the question is that do the diffusion constants also diverges in the $d\rightarrow \theta$ limit? We are going to study this special condition in details.

 Last but not least, in some of the previous literature \cite{blake,kim17}, the relation $D\sim v^2_B\tau$  was studied with non-zero dynamical exponent $z$ and hyperscaling violating factor $\theta$. But  only  one gauge field was considered in the action.  Actually, to realize Lifshitz geometry with $z> 1$, one needs to introduce at least one auxiliary gauge field in additional to the real Maxwell field \cite{dawei,tarrio}. The auxiliary gauge field is responsible for supporting the Lifshitz-like vacuum of the background, while the Maxwell field makes the black hole charged. It is interesting to check the relation $D\sim v^2_B\tau$ in the presence of two gauge fields.  Moreover, it was found in \cite{sonner} that finite DC electric conductivity can be realized simply because of the presence of the auxiliary $U(1)$ charge even without translational symmetry breaking. Another purpose of this paper is to examine the diffusion-butterfly velocity relation with such finite DC conductivity but without translational symmetry breaking.

  As a byproduct, we will also verify the universal formula of  dc electric conductivity proposed in \cite{unige17} for translational-symmetry broken BTZ-like black holes. This formula (i.e. $\prod_{i}\sigma_{ii}|_{q_i=0}=\prod_{i}Z^{d}_i \mathcal{A}^{d-2}$) states that the ratio of the determinant of the dc electrical conductivities along any spatial directions to the black hole area density $\mathcal{A}$ in zero-charge limit has a universal value.

 The structure of this paper is organized as follows. In section 2, we present the general formalism for the black hole solutions. Five concrete examples of BTZ-like black holes are derived. We briefly address dimensional reduction of BTZ black holes and the Jackiw-Teitelboim theory  in section 3. The diffusivity and butterfly-velocity  of charged BTZ black hole with momentum dissipation are discussed in section 4.  The transport properties of BTZ-like black holes with an extra hyperscaling violating factor are presented in section 5.  Discussions and conclusions are presented in section 6. In the Appendix, we provide  DC transport coefficients for general $d+2$-dimensional black holes in the presence of two $U(1)$ gauge fields.

\section{{The general formalism}}
 In order to show how the BTZ-like black hole solution emerges, we first consider a general $(d+2)$-dimensional action with an arbitrary Lifshitz dynamical exponent $z$ and a hyperscaling violating factor $\theta$
\begin{equation}\label{action1}
S=-\frac{1}{16\pi G_{d+2}}\int d^{d+2} x \sqrt{-g}[R+V(\phi)-\frac{1}{2}(\partial\phi)^2-\frac{1}{4}\sum_{i=1}^n Z_i(\phi) F_{(i)}^2-\frac{1}{2}Y(\phi)\sum\limits_{i}^{d}(\partial \chi_i)^2],
\end{equation}
where we will use the notation $Z_i= e^{\lambda_i \phi}$ and $Y(\phi)= e^{-\lambda_2\phi}$ in what follows and $\chi_i=\beta \delta_{Ii} x^i$ is a collection of $d-$massless linear axions introduced to break the translational symmetry and $\beta$ denotes the strength of momentum relaxation and disorder of the dual condensed matters.
 The action consists of Einstein gravity, axion fields, and $U(1)$ gauge fields and  a dilaton field. For simplicity, we only consider two $U(1)$ gauge $F^{(1)}_{rt}$ and $F^{(2)}_{rt}$ in which the first gauge field plays the role of an auxiliary field, making the geometry asymptotic Lifshitz, and the second gauge field is the exact Maxwell field making the black hole charged.

 The Einstein equation is given as
 \be
 R_{\mu\nu}=\frac{1}{2}\partial_{\mu}\phi\partial_{\nu}\phi+\frac{Y}{2}\sum_i \partial_{\mu}\chi_i\partial_{\nu}\chi_i+\frac{1}{2}\sum_i Z_i F_{(i)\mu}^{\rho}F_{\mu \rho}^{(i)}
 -\frac{g_{\mu\nu}}{4(d-2)}\sum_i Z_i F^2_{(i)}-\frac{V(\phi)}{d-2}g_{\mu\nu}.\nonumber
 \ee
 The equations of motion for the dilaton field and axion fields are obtained as
 \bea
 0&=&\Box\phi+V'(\phi)-\frac{1}{4}\sum_i Z'_i(\phi) F^2_{(i)}-\frac{1}{2}Y'(\phi)\sum_i\partial (\chi_i)^2,\\
 0&=&\nabla_{\mu}\bigg(Y(\phi)\nabla^{\mu}\chi_{i}\bigg).
 \eea
 The Maxwell field equation is
 \be
 0=\nabla_{\mu}\bigg(Z_i(\phi)  F^{\mu\nu}_{(i)}\bigg).
 \ee
 The dilaton field can be solved from the combinations of the (r,r) and (t,t) components  (i.e. $R^r_r-R^t_t$) of the Einstein equation and the solution read
 \be
 \phi=\nu\ln r=\sqrt{2(d-\theta)(z-1-\theta/d)}\ln r.
 \ee
 Assumed the metric takes the form $ds^2=r^{2\alpha}\bigg(-r^{2z}f(r)dt^2+dr^2/r^2 f(r)+r^2 d\vec{x}^2_{d}\bigg)$ and $\alpha=-\theta/d$,
 the Maxwell equation can be solved as
 \be
 F_{(i)rt}=Z^{-1}_i(\phi)r^{\alpha(2-d)+z-d-1}Q_i.
 \ee
 The metric function $f(r)$ can be solved from the $(x,x)$-component of the Einstein equation
 \bea\label{odef}
 \bigg[r^{d(\alpha+1)+z}f(r)\bigg]'=\frac{r^{\alpha(d+2)+z+d-1}}{(\alpha+1)}\bigg(\frac{V_0 r^{\gamma \nu}}{d}-\frac{1}{2d}\sum_{i}r^{-2d(\alpha+1)-\lambda_i \nu}Q^2_i-\frac{\beta^2}{2}r^{-\lambda_2\nu-2-2\alpha}\bigg).
 \eea
 The solution yields its form
 \bea
 f(r)&=&\frac{V_0 r^{2\alpha+\gamma \nu}}{d(\alpha+1)[\gamma \nu+\alpha(d+2)+z+d]}-\sum_i  \frac{Q_i^2 r^{-2\alpha(d-1)-\nu\lambda_i-2d}}{2d(\alpha+1)[\alpha(2-d)+z-d-\lambda_i \beta]}\nonumber\\
 &-& m r^{-d\alpha-z-d} -\frac{\beta^2_0 r^{-2z-2\alpha}}{2(1+\alpha)(d-z-2\alpha+d\alpha)},
 \eea
 where we have set $V(\phi)=V_0 r^{-2\alpha}$, $Z_i(\phi)=e^{\lambda_i \phi}$ and $Y(\phi)=1/Z_2$. We normalize the first term in the metric function to be one, so we have $\gamma=-2\alpha/\beta$ and $V_0=(d\alpha+z+d-1)(d\alpha+z+d)$.

 By further using the equation of motion for the dilaton field, we finally arrive at
 \be
 \bigg(\frac{4\nu}{d(\alpha+1)}-\frac{8\alpha}{\beta}\bigg)V=\sum_i  e^{\lambda_i \phi}F_{(i)}^2\bigg(\lambda_i-\frac{\nu}{d(\alpha+1)}\bigg)+ 2\alpha^2 \bigg(\frac{\nu}{\alpha+1}
 -\lambda_2 d\bigg)r^{-\lambda_{2}\nu-2\alpha}.
 \ee
 We then obtain the expressions for $\lambda_i$, $Q_1$ and $\alpha$
 \bea
 \lambda_1&=&-\frac{2\alpha(d-1)+2d}{\sqrt{2d(\alpha+1)(\alpha+z-1)}},~~~ \lambda_2=\sqrt{\frac{2(\alpha+z-1)}{d(\alpha+1)}},\\
 Q^2_1&=&\frac{2(z-1)V_0}{(d\alpha+z+d-1)}, ~~~ \alpha=-\theta/d.
 \eea
  The associated black hole solution was first obtained by some of us in \cite{ge16,ge1606,lv16}:
\bea\label{metric}
&& ds^2=r^{-\frac{2\theta}{d}}\bigg(-{{r^{2z}}}f(r)dt^2+\frac{dr^2}{r^2 f(r)}+{ {r^2}}d\vec{x}^2_{d}\bigg),\n\\
 &&f(r)=1-{\frac{m}{r^{d_{\theta}}}}+{\frac{Q^2_2}{r^{2(d_{\theta}-1)}}}-{\frac{ \beta^2}{r^{2z-2\theta/d}}},\label{fr}\\
 &&F_{(1)rt}=Q_1\sqrt{2(z-1)(d+z-\theta)}r^{d+z-\theta-1},\n\\
 &&F_{(2)rt}=Q_2 \sqrt{2(d-\theta)(d+z-\theta-2)}r^{-(d+z-\theta-1)},\n\\
 %&&\lambda_1=-\frac{2d-2\theta+\frac{2\theta}{d}}{\sqrt{2(d-\theta)(z-1-\theta/d)}}, ~~~
 %\lambda_2=-\sqrt{{2}{\frac{z-1-\theta/d}{d-\theta}}},\n\\
  %&&e^{\phi}=r^{\sqrt{2(d-\theta)(z-1-\theta/d)}},~~~V(\phi)=(d_{\theta}-1)d_{\theta}r^{-2\theta/d},~~ d_{\theta}=d+z-\theta,\n\\
  && \chi_i= \beta_{ia} x^a, ~~~\beta^2_0=\frac{1}{d}\sum^{d}_{i=1}\overrightarrow{\beta}_a \cdot \overrightarrow{\beta}_a,~~~\overrightarrow{\beta}_a \cdot \overrightarrow{\beta}_b= \beta^2_0 \delta_{ab} ~~~{\rm for}~~~ i\in \{1,d\}, \n
 \eea
 where $\beta^2={\frac{d^2 \beta^2_0}{2(d-\theta)(d^2+2\theta-(z+\theta)d)}}$.
% The chemical potentials and the charge density are given by

%\bea
%\mu_1&=&Q_1\bigg({\frac{2(z-1)}{d+z-\theta}}\bigg)^{1/2},~~~\mu_2=Q_2 \bigg({\frac{2(d-\theta)}{d+z-\theta-2}}\bigg)^{1/2},\\
%\rho_1&=& \mu_1 \rh^{-(d+z-\theta)},~~~~~~\rho_2=\mu_2 \rh^{d+z-\theta-2}.
%\eea
The above metric function holds under the condition that $Q_2$ and $\beta^2$ are finite and $Q_1$ does not contribute to the metric function, since $F_{(1)rt}$ is introduced to realizing the Lifshitz scaling. The constraints from the null energy condition of the gravity yields $(d-\theta)[d(z-1)-\theta]\geq 0$ and $(z-1)(d+z-\theta)\geq 0$.
  We recover the normal AdS geometry, when $F_{(1)rt}=0$ as the dynamical exponents take the values $z=1$ and $\theta=0$. The black hole solution can also recover that of the Lifshitz black hole solution as $\beta=0$ \cite{tarrio}.
  The event horizon locates at $r=\rh$ satisfying the relation $f(\rh)=0$. One can express the mass parameter $m$ in terms of $\rh$:
\be
m=\rh^{d+z-\theta}+Q^2_2 \rh^{2-d-z+\theta}-\beta^2 \rh^{d-z-\theta+2\theta/d}.
\ee
By further introducing an coordinate $u=\rh/r$, we can recast $f(r)$ as
\be\label{metric}
f(u)=1-u^{d+z-\theta}+\frac{Q^2_2}{\rh^{2(d+z-\theta-1)}}\bigg[u^{2(d+z-\theta-1)}-u^{d+z-\theta}\bigg]
+\frac{\beta^2}{\rh^{2z-2\theta/d}}\bigg[u^{d+z-\theta}-u^{2z-2\theta/d}\bigg].
\ee
The corresponding Hawking temperature is given by
\be\label{tem}
T=\frac{(d+z-\theta)\rh^z}{4\pi}\bigg[1-\frac{d+z-\theta-2}{d+z-\theta}Q^2_2\rh^{-2(d+z-\theta-1)}-\frac{d^2+2\theta-(z+\theta)d}{d (d+z-\theta)}\rh^{2\theta/d-2z}\beta^2\bigg].
\ee
The entropy density is given by $s=4 \pi\rh^{d-\theta}$. The specific heat of this black hole can be evaluated via $c_{Q_i,\beta}=T(\partial s/\partial T)_{Q_i,\beta}$.
The butterfly velocity is
\be
v^2_B=\frac{2\pi T \rh^{z-2}}{(d-\theta)}.
\ee
The DC transport coefficients in this general background are given in section 5. The more detail derivation of the DC transport coefficients in this background was obtained in  \cite{ge16,ge1606} (see also \cite{kuang17,lv17,bha,haishan,matteo} for more recent work).

We now present several examples when black hole solutions become ``critical"  as  the charge density $Q_2$ or axion density $\beta^2$ is formally zero. In other words, when one of the term in $f(r)$ is zero, we need to work out the solution very carefully because those terms would produce a  logarithmic term in the metric function. This can also be seen in the differential equation of $f(r)$, i. e. (\ref{odef}).   Such  a logarithmic term can greatly modify  the solution especially the behavior of the electric field.
The result is that the metric function $f$ \textit{in any $d\geq 1$ dimension } can be recast in a form similar to charged BTZ black holes in $2+1$ dimensions. In general, there are five conditions  that the logarithmic term appears in the metric function as summarized below:

\begin{table*}[htbp]\label{table}
\begin{center}
\begin{tabular}{|c|c|c|c|c|c|c|c|c|}
\hline
${\rm only} ~d+z-\theta-2=0$&$\rm the~ mass~  and~  Q_2 ~ related~  terms ~ are ~ degenerated $\\
\hline
${\rm only} ~d^2+2\theta-(z+\theta)d=0$&$\rm {the~ mass ~ and~  axion~  related~  terms ~ are~  degenerated}$\\
\hline
$ d=\theta,~ z\neq 2$&$\rm the~  Q_2~  and~  axion ~ related~  terms ~ are ~ degenerated$\\
\hline
$(d=\theta, z=2)~{\rm or}~(d=1, z=1+\theta) $&$\rm all~ mass,~Q_2~ and~ axions~ related~  terms ~ are ~ degenerated$\\

\hline
\end{tabular}
\caption{Summarization of five conditions  that the logarithmic term will appear in the metric function $f(r)$.}\label{table}
\end{center}
\end{table*}

$\bullet$ \textbf{Case I}: critical black hole solutions at $d+z-\theta-2=0$ and $d^2+2\theta-(z+\theta)d\neq 0$
Note that in this case, we have $d^2+2\theta-(z+\theta)d=2(2-z)(d-1)$. A well-defined solution can be achieved in the following form \cite{ge16,ge1606}
\bea
f(r)&=&1-{\frac{m}{r^{2}}}-\frac{q^2_2\ln r}{2(2-z)r^{2}}-{\frac{ \beta^2}{r^{2z-2\theta/d}}}\label{flog}\nonumber\\
&=& 1-\frac{\rh^2}{r^2}+\frac{q^2_2}{2r^2(2-z)}\ln \frac{\rh}{r}-\frac{\beta^2}{\rh^{2 z-{2 \theta }/{d}}}\bigg(\frac{\rh^2}{r^2}-\frac{\rh^{2 z-{2 \theta }/{d}}}{r^{2 z-{2 \theta }/{d}}}\bigg).,\nonumber\\
F_{(1)rt}&=&q_1 r,~~~ F_{(2)rt}=q_2 r^{-1}\label{frt}
\eea
where $m$, $q_1$ and $q_2$ are finite physical parameters without divergence as $(d+z-\theta-2)\rightarrow 0$.
 The metric function can recover that of charged BTZ black hole solution as $\beta=0$. A careful examination of (\ref{flog}) reveals that they satisfy the corresponding Einstein equation and Maxwell equation.
The Hawking temperature is given by
\be\label{tone}
T=\frac{\rh^z}{2\pi}\bigg(1-\frac{q^2_2}{4(2-z)\rh^2}-\frac{\beta^2 (d+\theta-dz)}{d \rh^{2 z-{2 \theta }/{d}}}\bigg).
\ee

$\bullet$ \textbf{Case II}: critical black hole solutions at  $d+z-\theta-2\neq 0$ and $d^2+2\theta-(z+\theta)d= 0$
The metric function and gauge fields in this case take their forms
\bea
&& f(r)=1-\frac{m}{r^{d+z-\theta}}+\frac{q^2_2}{r^{2(d+z-\theta-1)}}-\frac{d\beta^2_0}{2(d-\theta)r^{d+z-\theta}}\ln r,\\
&&F_{(1)rt}=q_1r^{d+z-\theta-1},~~~
 F_{(2)rt}=q_2 r^{-(d+z-\theta-1)}.
\eea
We can also express the metric function in terms of the event horizon radius
\be
f(r)=1-\bigg(\frac{\rh}{r}\bigg)^{d+z-\theta}+\frac{q^2_2}{\rh^{2(d+z-\theta-1)}}\bigg[\bigg(\frac{\rh}{r}\bigg)^{2(d+z-\theta-1)}-\bigg(\frac{\rh}{r}\bigg)^{d+z-\theta}\bigg]
-\frac{\beta^2}{r^{d+z-\theta}}\ln\frac{\rh}{r}.
\ee
Note that only the linear scalar term in the metric function becomes a logarithmic term.
The black hole temperature yields
\be
\label{ttwo}
T=\frac{\rh^{z}}{4\pi}\bigg[(d+z-\theta)-\left(\beta^2 \rh^{\theta-d-z}+q^2_2 \rh^{-2(d+z-\theta-1)}(d+z-\theta-2)\right)\bigg].
\ee
On the other hand, as $q_2\rightarrow 0$  the metric is analogous to the charged BTZ black hole metric.

$\bullet$~\textbf{Case III}:
As $d=\theta$ but $z\neq 2$, the two terms containing $Q_{2}^{2}$ and $\beta^{2}$ in the metric function degenrate, that is to say
\begin{eqnarray*}
\frac{Q_{2}^{2}}{r^{2(d+z-\theta-1)}}-\frac{\beta^{2}}{r^{2(d+z-\theta-1)}}r^{\frac{2}{d}(d-\theta)(d-1)} & \to & \frac{Q_{2}^{2}-\beta^{2}}{r^{2(d+z-\theta-1)}}-\frac{2}{d}(d-\theta)(d-1)\beta^{2}\frac{\ln r}{r^{2(d+z-\theta-1)}}\\
 & = & \frac{Q_{2}^{2}-\beta^{2}}{r^{2(d+z-\theta-1)}}-\frac{(d-1)\beta_{0}^{2}}{d+\frac{2\theta}{d}-z-\theta}\frac{\ln r}{r^{2(d+z-\theta-1)}}\\
 & = & \frac{q^{2}_2}{r^{2(d+z-\theta-1)}}-\frac{(d-1)\beta_{0}^{2}}{d+\frac{2\theta}{d}-z-\theta}\frac{\ln r}{r^{2(d+z-\theta-1)}},
\end{eqnarray*}
where we have used the relation
\be r^{\frac{2}{d}(d-\theta)(d-1)}\to1+\frac{2}{d}(d-\theta)(d-1)\ln r \ee and  $q^{2}_2:=Q_{2}^{2}+\beta^{2}$. So  in this condition, the metric function can be written as
\begin{eqnarray*}
f(r) & = & 1-\frac{m}{r^{d+z-\theta}}+\frac{q^{2}_2}{r^{2(d+z-\theta-1)}}-\frac{(d-1)\beta_{0}^{2}}{d+\frac{2\theta}{d}-z-\theta}\frac{\ln r}{r^{2(d+z-\theta-1)}}\\
 & = & 1-\frac{m}{r^{z}}+\frac{q^{2}_2}{r^{2(z-1)}}-\frac{(d-1)\beta_{0}^{2}}{2-z}\frac{\ln r}{r^{2(z-1)}},\\
 F_{(1)rt}& = &q_1r^{z-1},~~~
 F_{(2)rt}=q_2 r^{-(z-1)}.
\end{eqnarray*}
Here $q_2$ and $\beta_{0}$ are finite and  physical parameters
under the $d-\theta\to0$ limit ($z\ne2$), instead of  diverging $Q_{2}$ and
$\beta$. The metric function written in terms of the event horizon is given by
\be
f(r)=1-\bigg(\frac{\rh}{r}\bigg)^z+\frac{q^2_2}{r^{2z-2}}\bigg[1-\bigg(\frac{\rh}{r}\bigg)^{2-z}\bigg]+\frac{(d-1)\beta^2_0}{(z-2)r^{2z-2}}\bigg((\frac{\rh}{r})^{2-z}\ln \rh-\ln r\bigg).
\ee
The Hawking temperature evaluated at the event horizon yields
\be
T=\frac{z \rh^z}{4\pi}-\frac{q^2_2(z-2)\rh^{2-z}}{4\pi}+\frac{(d-1)\beta^2_0\rh^{2-z}}{4\pi(z-2)}\bigg[1-(z-2)\ln\rh\bigg].
\ee
%\[
%f(r_{H})=0\Longrightarrow
%\]
%\[
%M=r_{H}^{z}+Q^{2}r_{H}^{-(z-2)}-\frac{(d-1)\beta_{0}^{2}}{2-z}r_{H}^{-(z-2)}\ln r_{H}
%\]
%\[
%f(r)=1-\frac{r_{H}^{z}}{r^{z}}+\frac{Q^{2}}{r^{z}}(r^{-(z-2)}-r_{H}^{-(z-2)})-\frac{(d-1)\beta_{0}^{2}}{(2-z)r^{z}}(r^{-(z-2)}\ln r-r_{H}^{-(z-2)}\ln r_{H})
%\]
%\begin{eqnarray*}
%f^{\prime}(r) & = & \frac{z}{r}\frac{r_{H}^{z}}{r^{z}}-\frac{Q^{2}}{r^{z+1}}(2\frac{z-1}{r^{z-2}}-\frac{z}{r_{H}^{z-2}})-\cdots(r^{-(z-2)}\ln r-r_{H}^{-(z-2)}\ln r_{H})\\
% &  & -\frac{(d-1)\beta_{0}^{2}}{(2-z)r^{z}}(\frac{1}{r^{z-1}}-\frac{z-2}{r^{z-1}}\ln r)
%\end{eqnarray*}
%\[
%f^{\prime}(r_{H})=\frac{z}{r_{H}}-\frac{(z-2)Q^{2}}{r_{H}^{2z-1}}-\frac{(d-1)\beta_{0}^{2}}{(2-z)r_{H}^{2z-1}}[1-(z-2)\ln r_{H}]
%\]

$\bullet$~\textbf{Case IV}: critical black hole solutions at $d+z-\theta-2= 0$ and $d^2+2\theta-(z+\theta)d= 0$
This condition yields two solutions $d=1,~~ z=1+\theta$ or $d=\theta,~~ z=2$.  We focus on the $d=1$ and $z= 1+\theta$ here and obtain a very particular
black hole solution
\bea \label{hyperbtz}
ds^2&=&r^{-2\theta}\bigg(-r^{2z}f(r)dt^2+\frac{dr^2}{r^2f(r)}+r^2 dx^2\bigg),\\
f(r)&=&1-\frac{m}{r^2}-\frac{q^2_2+\beta^2_0}{2(1-\theta)r^2}\ln r\\
&=& 1-\frac{\rh^2}{r^2}+\frac{q^2_2+\beta^2_0}{2(1-\theta)r^2}\ln \frac{\rh}{r},\\
F_{(1)rt}&=&q_1 r, ~~~F_{(2)rt}=q_2 r^{-1}.
\eea
The scalar potentials are given by \be A_{(1)t}=\mu_1\bigg(1-\frac{r^2}{\rh^2}\bigg),~~~A_{(2)t}=q_2 \ln \frac{r}{\rh}.
\ee
The chemical potentials and charge density are given by $\mu_1$ and $\mu_2=- q_2\ln\rh$, respectively. Notice that $\mu_1$ does not correspond to the chemical potential of the black hole and as $r\rightarrow \infty$, $A_{(1)t}\rightarrow \infty$.
The event horizon locates at $r=\rh$ satisfying $f(\rh)=0$. The Hawking temperature is given by
\be
T=\frac{\rh^{1+\theta}}{2\pi}\bigg(1-\frac{q^2_2+\beta^2_0}{4(1-\theta)\rh^2}\bigg).
\ee
The metric function is actually a three-dimensional BTZ-like black hole solution with a Lifshitz dynamical exponent and a hyperscaling factor.
As $\theta=0$ and then $q_1=0$, we recover the standard charged BTZ black hole solution obtained in (\ref{btz}) since $\theta=0$ and $d+z-\theta=2$ infers $z=1$.
The entropy density and the butterfly velocity are
\be
s=\frac{\rh^{1-\theta}}{4G}, ~~~v^2_B=\frac{2\pi T \rh^{z-2}}{(d-\theta)}.
\ee
The relation between the butterfly velocity and diffusion constant will be studied in detail in what follows.

$\bullet$ \textbf{Case V}:  As to $d=\theta$ and $z=2$, the null energy condition is satisfied. The metric becomes
\bea
ds^2&=&-r^2 f(r)dt^2+\frac{dr^2}{r^4 f(r)}+d\vec{x}^2,\\
f(r)&=& 1-\frac{\rh^2}{r^2}+\frac{(q^2_2+\beta^2_0)}{r^2}\ln \frac{\rh}{r},\\
F_{(1)rt}&=&q_1 r, ~~~F_{(2)rt}=q_2 r^{-1}.
\eea
The corresponding Hawking temperature is $T=\frac{\rh}{\pi}(1-\frac{q^2_2+\beta^2_0}{4\rh^2})$. The entropy density is a constant $s=1/4G$. The metric given here is very special since it has non-zero Hawking temperature, but constant entropy density. The associated specific heats are also vanishing. The butterfly velocity becomes divergent as $d \rightarrow \theta$. The transport coefficients in this case are also very special
\bea
\sigma_{11}&=&\frac{1}{\rh^{2}}+\frac{q^2_1}{\beta^2_0 },~~~ \sigma_{12}=\sigma_{21}=\frac{q_1 q_2}{\beta^2_0 },\\
\sigma_{22}&=&1+\frac{q^2_2}{\beta^2_0  }, ~~~\alpha_1=\frac{4\pi q_1}{\beta^2_0}, \alpha_2=\frac{4\pi q_2}{\beta^2_0},\\
\bar{\kappa}&=&\frac{16 \pi^2 T  }{\beta^2_0}
\eea
Both dc electric conductivity  $\sigma_{22}$ driven by the real Maxwell field and the thermoelectric conductivities $\alpha_1$ and $\alpha_2$ are constants.  It is worth future investigating whether this black hole solution and its boundary dual has any more  physical meaning.

In the following discussions, we first discuss the relation between the charged BTZ black hole and its dimensioanl reduction to Jackiw-Teitelboim theory.
Then we study diffusions and butterfly velocity  for momentum dissipated-charged BTZ black holes. The hyperscaling violating factor modified BTZ black holes will also be examined.

\section{Dimensional reduction and Jackiw-Teitelboim theory}
The gravitational duality of the original SYK model is argued to be a two-dimensional dilaton $AdS_2$ gravity \cite{jensen}, i.e. the Jackiw-Teitelboim theory. A three-dimensional gravity can be reduced to Jackiw-Teitelboim theory by dimensional reduction. In the context of the $AdS_3/CFT_2$ duality, a two-dimensional CFT has a holographic bulk dual. At the high temperature the thermal state of the theory is dual to a BTZ black hole.  In
Consider a three-dimensional gravity in general
\begin{equation}\label{3D}
S=-\frac{1}{16\pi G_{3}}\int d^{3} x \sqrt{-g}[R+2 \Lambda-\frac{1}{4} F^2],
\end{equation}
The corresponding black hole solution is the BTZ black hole solution of  the form
\bea
ds^2&=&-{{r^{2}}}f(r)dt^2+\frac{dr^2}{r^2 f(r)}+{ {r^2}}d{x}^2,\\
f(r)&=&1-\frac{\rh^2}{r^2}+\frac{q^2}{2r^2}\ln\frac{\rh}{r}, ~~~F_{rt}=-\frac{q}{r}.
\eea
Assume that there is a single coordinate called $\varphi$, which is independent of the gravitational field in three dimensions and the metric takes the form
\be\label{reduction}
ds^2=h_{ij}(x^i)dx^i dx^j+\Phi^2(x^i)d\varphi^2,
\ee
where $i,j=0,1$.
The action (\ref{3D}) simply reduces to  that of the Jackiw-Teitelboim theory
\be
S=\int d^2 x \sqrt{-h}~\Phi [R+2 \Lambda -\frac{1}{4} F^2].
\ee
The  BTZ solutions written in the form as (\ref{reduction}) yield a solution to the Jackiw-Teitelboim model
\bea
ds^2&=&-{{r^{2}}}f(r)dt^2+\frac{dr^2}{r^2 f(r)},~~~\Phi=r,\\
f(r)&=&1-\frac{\rh^2}{r^2}+\frac{q^2}{2r^2}\ln\frac{\rh}{r}, ~~~F_{rt}=-\frac{q}{r}.
\eea
In \cite{wenhe}, some of us consider  a new higher dimensional SYK model with complex fermions on bipartite lattices and obtain linear in temperature resistivity, thermal conductivity and specific heat.

\section{Diffusion and butterfly velocity of disordered BTZ black holes }
\begin{figure}
\begin{center}
	\includegraphics[height=7cm,trim={3cm  18cm 5cm 2.7cm},clip]{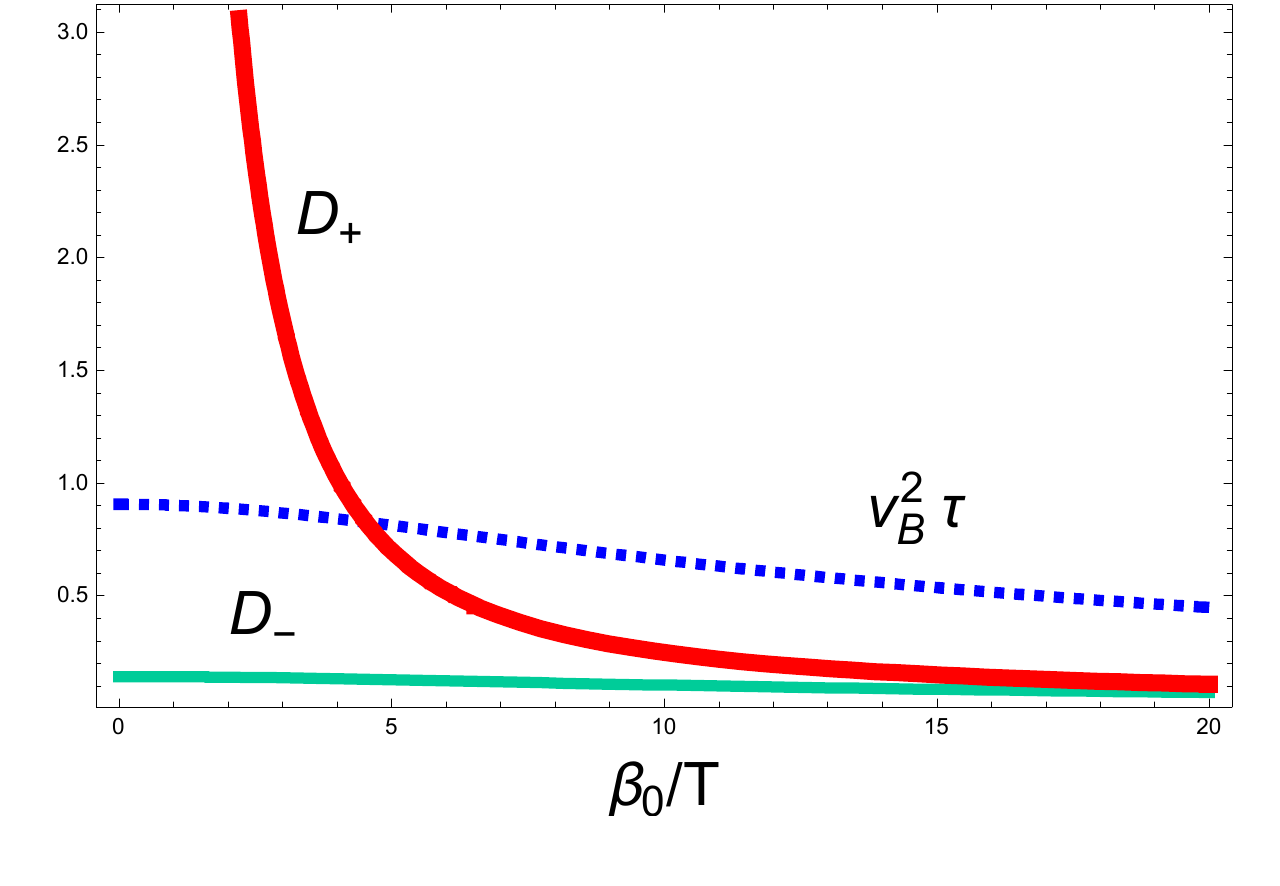}
	\includegraphics[height=6.5cm,trim={3cm  19cm 5cm 2.7cm},clip]{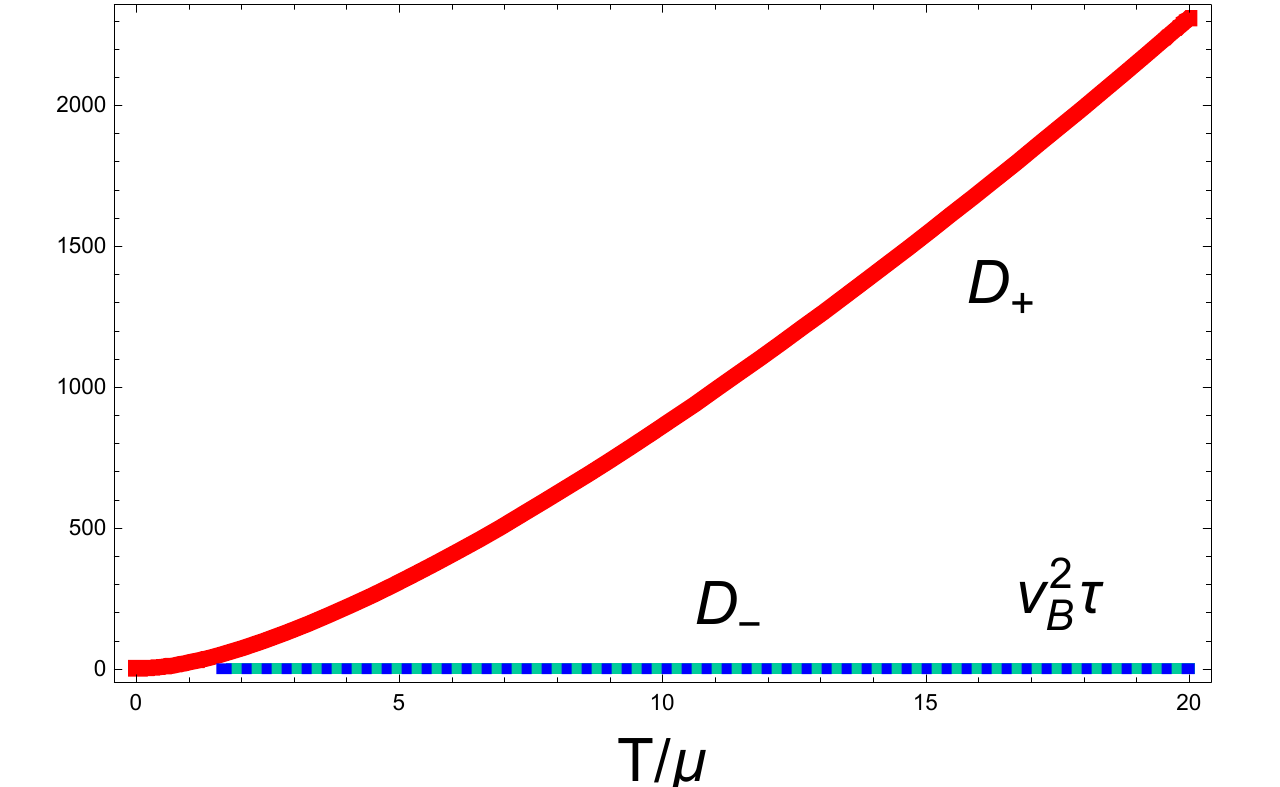}
\end{center}
\caption{Diffusion constants $D_{\pm}$ and the butterfly effect $v^2_{B}\tau$ as functions of the axion field and temperature. (Top) The diffusion constants $D_{+}$ (red), $D_{-}$ (green) and $v^2_B\tau$ (blue) as functions of $\beta_0$, where we have set $\mu=1$ and $T=1$. (Bottom) The diffusion constants $D_{+}$, $D_{-}$ and $v^2_B\tau$ as functions of $\beta_0$, where we have set $\mu=1$ and $\beta_0=1$.}\label{btz12}
\end{figure}
A special case is $d=1$, $\phi=0$, $V(\phi)=2$, $Y(\phi)=1$ and $Z(\phi)=1$. This corresponds to the Case IV discussed in section 2. In this case, the action reduces to that of BTZ black hole with a momentum dissipation term:
\bea
S=-\frac{1}{16\pi G_{3}}\int d^{3} x \sqrt{-g}[R+2 \Lambda-\frac{1}{2}(\partial\chi)^2-\frac{1}{4} F^2]-\frac{1}{8\pi G_{3}}\int_{\partial M}d^2 x\sqrt{-h}K+S_{\rm ct},\nonumber
\eea
It is well known that the naive free energy of the charged BTZ black hole is logarithmically divergent. Jensen found the divergence is due to the Weyl anomaly of the boundary CFT \cite{jensen}. The counter-term is given by \cite{jensen}
\be
S_{\rm ct}=\frac{1}{8\pi G_{3}}\int_{r=r_\Lambda}d^2 x\sqrt{-h}\bigg(1+\frac{R[h]}{2}-\frac{1}{4}\ln r_\Lambda \partial_i \chi \partial^i \chi \bigg)-\frac{1}{2}\int_{r=r_\Lambda}d^2 x\sqrt{-h}F_{ra}F^{ra}\ln r_\Lambda.\nonumber
\ee
One of the solutions to the above action is given by the $2+1$-dimensional charged BTZ metric and a linear scalar field
\bea \label{btz}
&&ds^2=\frac{1}{z^2}\bigg(-f(z)dt^2+dx^2+\frac{dz^2}{f(z)}\bigg),\n\\
&&f(z)=1-\frac{z^2}{z^2_h}+\frac{q^2+\beta^2_0}{2}z^2\ln\frac{z}{z_h},\\
&&A_t(z)=q \ln\frac{z}{z_h},~~~\chi_1=\beta_0 x.\n
\eea
The black hole temperature is given by  $ T= \frac{4-z^2_h(q^2+\beta^2_0)}{8\pi z_h}$ .
The charge density and chemical potential can be obtained from the near-boundary expansion of the gauge field $j^t=q$ and $\mu=-q\ln \rh$.
The renormalized free energy is given by
\be
F=-\frac{1}{2}Ts+\frac{1}{2}\mu q+\frac{1}{4}q^2+\beta_0 \Phi,
\ee
where $\Phi=-\frac{1}{2}\beta_0 \ln(2\pi T+\sqrt{q^2_1+\beta^2_0+4\pi^2 T^2})$.
 The horizon radius expressed in terms of $T$, $q$ and $\beta_0$ goes as
\be
z_h=\frac{2}{\beta^2_0+q^2}\bigg(\sqrt{4\pi^2 T^2+\beta^2_0+q^2}-2\pi T\bigg).
\ee
The entropy density reads
\be
s=4\pi z^{-1}_h=4\pi \bigg(\frac{1}{2} \sqrt{\beta^2_0+q^2+4 \pi ^2 T^2}+\pi  T \bigg).
\ee
The butterfly velocity can be evaluated as
\be
v^2_B={2\pi T}z_h=\frac{4 \pi T}{\beta^2_0+q^2}\bigg(\sqrt{4\pi^2 T^2+\beta^2_0+q^2}-2\pi T\bigg).
\ee
 The DC conductivity is obtained as
\be\label{btzs}
\sigma={z_h}+\frac{ q^2}{\beta^2_0 }z_h=\frac{2}{\beta^2_0} \bigg(\sqrt{4\pi^2 T^2+\beta^2_0+q^2}-2\pi T \bigg).
\ee
The compressibility and the thermoelectric susceptibility can be expressed as
\bea
\chi&=& \bigg(\frac{\partial q}{\partial \mu}\bigg)_{T}=\frac{-{a_{11}}}{{a_{11}} \log \left(\frac{1}{2} \sqrt{{\beta_0}^2+q^2+4 \pi ^2 T^2}+\pi  T\right)+q^2},\\
\zeta&=&\bigg(\frac{\partial s}{\partial \mu}\bigg)_{T}=\frac{-2 \pi  q \left(\sqrt{\beta_0^2+q^2+4 \pi ^2 T^2}+2 \pi  T\right)}{a_{11} \log \left(\frac{1}{2} \sqrt{\beta_0^2+q^2+4 \pi ^2 T^2}+\pi  T\right)+q^2},
\eea
where $a_{11}= \beta_0^2 + q^2 +
  2 \pi T (2 \pi T + \sqrt{\beta_0^2 + q^2 + 4 \pi^2 T^2})$.
The specific heat at fixed charge density are
\bea
c_{q}&=&T\bigg(\frac{\partial s}{\partial T}\bigg)_{q}=4 \pi^2 T+\frac{8 \pi ^3 T^2}{\sqrt{{\beta_0}^2+q^2+4 \pi ^2 T^2}}.
\eea
The temperature dependence of the entropy density and the specific heat obtained here is similar to that of $1+1$-dimensional SYK model with complex fermions on bipartite
lattices \cite{wenhe}.
The Seebeck coefficient and the thermal conductivity are given by
\bea
\alpha &=&\bar{\alpha}=\frac{4\pi q}{\beta^2_0}, \\
\bar{\kappa}&=&\frac{16 \pi^2 T }{z_h \beta^2_0}=\frac{16 \pi^2 T }{\beta^2_0}\bigg(\frac{1}{2} \sqrt{\beta^2_0+q^2+4 \pi ^2 T^2}+\pi  T \bigg),\\
\kappa&=&\bar{\kappa}-\frac{\alpha^2 T}{\sigma}=\frac{32 \pi ^2 T \left(\left(\frac{1}{2} \sqrt{\beta^2_0+q^2+4 \pi ^2 T^2}+\pi  T\right)^2-\frac{q^2}{{\beta_0}^2+q^2}\right)}{{c}^2_1 \left(\sqrt{\beta^2_0+q^2+4 \pi ^2 T^2}+2 \pi  T\right)}.
\eea
In general, the diffusion constants $D_{\pm}$ are related to the transport coefficients and thermal quantities through the Einstein relations \cite{hartnoll14}
\bea\label{dif}
D_{+}D_{-}&=&\frac{\sigma}{\chi}\frac{\kappa}{c_{\rho}},\\
D_{+}+D_{-}&=&\frac{\sigma}{\chi}+\frac{\kappa}{c_{\rho}}+\frac{T(\xi\sigma-\chi \alpha)^2}{c_{\rho}\chi^2\sigma}.
\eea
The diffusion constants are solved as
\be
D_{\pm}=\frac{c_2\pm\sqrt{c^2_2-4c_3}}{2},
\ee
where
\be\label{58}
c_2=\frac{\sigma}{\chi}+\frac{\kappa}{c_{\rho}}+\mathcal{M},~~~ c_3=\frac{\sigma}{\chi}\frac{\kappa}{c_{\rho}},~~~ \mathcal{M}=\frac{T(\zeta \sigma-\chi \alpha)^2}{c_{\rho}\chi^2\sigma}.
\ee
In the absence of the mixing term $\mathcal{M}$, $D_{+}$ and $D_{-}$ simply reduce to the charge diffusion $D_{c}$ and the energy diffusion $D_{e}$.
We then find the two diffusion constants in the strong momentum dissipation limit ($\beta_0/T\gg 1$) turn out to be
\bea
D_{+}&=&\frac{2}{\beta_0}-\frac{4\pi T}{\beta^2_0}+...,\\
D_{-}&=&\frac{2}{\beta_0}-\frac{4\pi T}{\beta^2_0}+...
\eea
The mixing term $\mathcal{M}$ has similar temperature dependence as $D_{-}$
\be
\mathcal{M}=\frac{8 q^2}{ \beta^3_0}+...
\ee
Compared with $D_{-}$, the mixing term $\mathcal{M}$ can be neglected in the incoherent regime.
The term $v^2_B \tau$ expanded at large $\beta_0$ yields
\be
v^2_B \tau=\frac{2\pi }{\rh }=\frac{4\pi}{\beta_0}-\frac{8\pi^2 T}{\beta^2_0}+...
\ee
We can see that in the incoherent regime $\beta_0/T\gg 1$, $D_{\pm} \sim v^2_B \tau$ with $\tau=\hbar/k_B T$. This implies that the BTZ black hole indeed satisfies the proposed relation between the diffusion constant and the butterfly velocity. As shown in Fig.\ref{btz12}, the diffusion constant $D_{\pm}$ approximate to a value parallel to the line $v^2_B \tau$  in the incoherent regime  when the temperature is fixed. However, as the axion $\beta_0/\mu$ is fixed,   $D_{-}$ coincides with $v^2_B \tau$ and $D_{+}\geq v^2_B \tau$ as the temperature goes up.

The low temperature and low charge density expansion of the diffusion constants and butterfly velocity is as follows
\bea
D_{+}&=&\frac{2}{\beta_0}-\frac{2\pi T}{\beta^2_0}+\mathcal{O}(T^2),\label{66}\\
D_{-}&=&\frac{2}{\beta_0}-\frac{2\pi T}{\beta^2_0}+\mathcal{O}(T^2),\label{67}\\
v^2_B \tau&=& \frac{4\pi}{\sqrt{\beta^2_0+q^2}}-\frac{8\pi^2 T}{\beta^2_0+q^2}+\mathcal{O}(T^2).
\eea
The mixing term between the charge- and energy- diffusion is given as
\be
\mathcal{M}=\frac{2 q^2 }{\beta_0^2 \sqrt{\beta_0^2+q^2}}\left(1-\frac{\beta_0^2+q^2}{\left(\beta_0^2+q^2\right) \ln \left(\frac{1}{2} \sqrt{\beta_0^2+q^2}\right)+q^2}\right)^2+\mathcal{O}(T),
\ee
which cannot be ingored at low temperature but finite charge density so $D_{+}$ and $D_{-}$ may not interpreted as charge diffusion $D_{c}$ and energy diffusion $D_{e}$.  As shown in Fig.\ref{btz12}, both in the incoherent regime and in the low temperature limit, $v^2_B \tau$ has the trend to go to zero, implying there is no propagation of the chaos in these cases.
One can also check the dimension of these quantities here:
$[q]=[\beta_0]=1, [\alpha]=-1, [r]=[1/z]=1$.    Assuming $ [T]=1$,  we have
$[\sigma]=-1,  [\kappa]=-1, $    $[\alpha^2 T/\sigma ]=-1$,  $ [D_{\pm}]=-1$  and $ [v^2_B \tau]=-1$  which is consistent. Notice that in \cite{wenhe}, $1+1$-dimensional SYK model with complex fermions on bipartite lattices  shows similar temperature dependence of the electric conductivity  and the charge diffusion.

On the other hand, we are able to check the validity of the universal dc electric conductivity  proposed in \cite{unige17} \be \label{scaling}
\frac{\prod_{i} {\sigma_{ii}}}{ \mathcal{A}^{d-2}}\bigg|_{q_i=0} = \prod_{i} Z_i^{d}\bigg|_{r=\rh}.
\ee
Substituting equation (\ref{btzs}) at zero charge density, the black hole area and gauge coupling $Z_2=1$ into (\ref{scaling}) and evaluating at the event horizon, we can see that (\ref{scaling}) is satisfied.

\section{Diffusion of BTZ-like black holes with a hyperscaling violating factor }
In order to check the universal relations between diffusivity and the butterfly velocity, let us first work with arbitrary parameters $d$, $z$ and $\theta$. After obtaining the general ansatz for $D_{+}$ and $D_{-}$, we then discuss the five black hole solutions given in section 2 and check the diffusion-butterfly effect relation. For simplicity of calculation, we mainly
consider the incoherent limit as follows
\be
\frac{\beta_0}{\mu}\gg 1, ~~~\frac{\beta_0}{T}\gg 1, {\rm with} ~~\frac{\mu}{T }~~~ {\rm finite}.
\ee
In addition, we only focus on the charge- and energy-diffusions and thus neglect the mixing term $\mathcal{M}$ in (\ref{58}).
The metric function and the Hawking temperature are given in (\ref{metric}) and (\ref{tem}), respectively. The specific heat at fixed charge density is given by
\be
c_q=T \bigg(\frac{\partial s}{\partial T}\bigg)_{q}=\frac{2 \partial_r g^{1/2}_{xx}}{\sqrt{g_{rr}}}\partial_r g^{d/2}_{xx}\bigg|_{\rh}\frac{\partial \rh}{\partial T}=4\pi T(d-\theta)\rh^{d-\theta-1}\frac{\partial \rh}{\partial T}.
\ee
The charge density yields
\be
q_i=\frac{g^{d/2}_{xx}}{\sqrt{g_{rr}g_{tt}}}Z_2 F_{(i)rt}.
\ee
The chemical potential is thus $\mu_2={q_2 \rh^{2-d-z+\theta}}/{(d+z-\theta-2)}$ and one can obtain the chemical potential of Reinssner-Nordstrom black hole $\mu=q_2/\rh$ as $d=2$, $z=1$ and $\theta=0$.
In general, the chemical potential depends on the details of the full bulk geometry. However, the black hole solution obtained here  mainly  describe the IR geometry. We assume that  the infra-red region of the geometry that dominates the behavior of the charge charge compressibility in what follows \cite{blake15}.
The compressibility is found to be
\bea
\chi&=& \bigg(\frac{\partial q_2}{\partial \mu}\bigg)_{T}=(2+\theta-d-z)\rh^{d+z-\theta-2}.
\eea
Since there are two $U(1)$ gauge fields, the DC electric conductivity in this case is actually a $2\times 2$ matrix. The general ansatz has been obtained by some of us in \cite{ge1606}
\bea
&&\sigma_{11}=\bigg(g^{\frac{d-2}{2}}_{xx}Z_1(\phi)+\frac{q^2_1}{\beta^2_0 Y(\phi)g^{d/2}_{xx}}\bigg)\bigg|_{r=\rh}=\rh^{2\theta-2\theta/d-2d}+\frac{q^2_1}{\beta^2_0}\rh^{2z+\theta-2-d-2\theta/d},\\
&&\sigma_{22}=\bigg(g^{\frac{d-2}{2}}_{xx}Z_2(\phi)+\frac{q^2_2}{\beta^2_0 Y(\phi)g^{d/2}_{xx}}\bigg)\bigg|_{r=\rh}=\rh^{d+2z-\theta-4}+\frac{q^2_2}{\beta^2_0}\rh^{2z+\theta-2-d-2\theta/d},\\
&&\sigma_{12}=\sigma_{21}=\frac{q_2 q_1}{\beta^2_0}\rh^{2z+\theta-2-d-2\theta/d}.
\eea
The thermoelectric conductivity $\alpha$ and thermal conductivity $\bar{\kappa}$ are obtained as
\bea
&&\bar{\alpha}_1=\frac{4\pi q_1}{\beta^2_0 Y(\phi)}\bigg|_{r=\rh}=\frac{4\pi q_1}{\beta^2_0 }\rh^{2z-2-2\theta/d},\\
&&\bar{\alpha}_2=\frac{4\pi q_2}{\beta^2_0 Y(\phi)}\bigg|_{r=\rh}=\frac{4\pi q_2}{\beta^2_0 }\rh^{2z-2-2\theta/d},\\
&&\bar{\kappa}=\frac{16 \pi^2 T g^{d/2}_{xx}}{\beta^2_0 Y(\phi)}\bigg|_{r=\rh}=\frac{16\pi^2 T}{\beta^2_0}\rh^{d+2z-2-\theta-2\theta/d}.
\eea
One may notice that even $q_1$ is zero now, $\sigma_{11}=\rh^{2\theta-2\theta/d-2d}$ has non-trivial contributions from the pair production of the boundary dual quantum field.
 That is to say, once the fluctuations of the auxiliary gauge field are turned on, there exists a discontinuity from the expressions of transport coefficients of Lifshitz spacetime to those of the AdS geometry.
 Assuming the charge current induced by the auxiliary gauge field is vanishing $J_1=0$, we are able to  focus on the diagonal elements \cite{ge16}
\bea
\sigma_{DC}&=&\rh^{d+2z-\theta-4}+\frac{q_2 \rh^{2\theta+2z-d-2\theta/d}}{\beta^2 \rh^{2+\theta}+q^2_1 \rh^{2z+d}},\nonumber\\
\alpha_{DC}&=&\bar{\alpha}_{DC}=\frac{4\pi \rh^{2z+\theta-2\theta/d}}{\beta^2 \rh^{2+\theta}+q^2_1 \rh^{2z+d}},\nonumber\\
\bar{\kappa}_{DC}&=&\frac{16\pi^2 T \rh^{2z+d-2\theta/d}}{\beta^2 \rh^{2+\theta}+q^2_1 \rh^{2z+d}}.\nonumber
\eea
Similarly, we have the thermal conductivity at zero electric current
\be
\kappa_{DC}=\frac{16 \pi^2 T \rh^{3d+2z}}{q^2_2 \rh^{4+3\theta}+(\beta^2 \rh^{2+\theta}+q^2_1 \rh^{d+2z})\rh^{2d+2\theta/d}}.
\ee
Considering only the charge diffusion $D_{c}$ and the energy diffusion $D_{e}$ and in the strong momentum relaxation limit, we obtain
\bea\label{dec}
D_{c}&=&\frac{\sigma}{\chi}=\frac{\rh^{z-2}}{d-\theta+z-2}+\mathcal{O}(\frac{1}{\beta^2_0}),\\
D_{e}&=&\frac{\kappa}{c_q}=\frac{\rh^{z-2} (d z-2 \theta ) \left(d^2-d \theta -d z+2 \theta \right)}{d^2 (d-\theta )}+\mathcal{O}(\frac{1}{\beta^2_0}).
\eea
Hereafter, our discussion work in the strong momentum relaxation limit. Singularities can be observed as $d+z-\theta=2$ and $d=\theta$. We may able to remove the singularities since the metric become critical in these cases.
The butterfly velocity can be computed by considering a shock wave geometry and written in terms of the metric at the horizon \cite{lingbutter,feng}
\be\label{butter}
v^2_{B}=\frac{4 \pi^2 T^2}{\sqrt{g_{xx}}m^2}\bigg|_{\rh},~~~ m^2=\pi T \frac{\partial_r g^d_{xx}}{g^d_{xx} \sqrt{-g_{rr}g_{tt}}}\bigg|_{\rh}.
\ee
In our case $v^2_{B}=\frac{2\pi T \rh^{z-2}}{(d-\theta)}$. From (\ref{dec}) and (\ref{butter}), we see that the relationship
\bea
D_{c}&=&\frac{(d-\theta)}{2 \pi(d-\theta+z-2)}v^2_B \tau,\\
D_{e}&=& \frac{ (dz-2\theta)(d^2-d\theta-dz+2\theta)}{2 \pi d^2}v^2_B \tau.
\eea
holds independently of the details of the bulk solution. However, from the expressions of the diffusion constants (\ref{dec}), we cannot obtain the diffusion constants of BTZ black holes evaluated in section 4. So do the five examples listed in table \ref{table}.
 The reasons are that the black hole temperature and the chemical potential are greatly modified for those cases.
 As $\theta\rightarrow 0$, The relations between diffusion constants and the butterfly velocity are given by
\bea
D_{c}&=&\frac{d}{2 \pi(d+z-2)}v^2_B \tau,\\
D_{e}&=& \frac{(d-z)z}{2 \pi}v^2_B \tau.
\eea
We discuss the diffusion constants of the five special cases listed in table \ref{table} in details:\\
$\bullet$ {Case I: $d-\theta+z-2=0$, $(d^2-d\theta-dz+2\theta)\neq 0$}\\
In this case, the chemical potential becomes logarithmic $\mu=- q \ln \rh$.  The Hawking temperature is reformulated in (\ref{tone}). The diffusion constants behave as
\bea\label{de}
D_{c}&=&\frac{\sigma}{\chi}=-{\rh^{z-2}} \ln \rh+\cdots,\\
D_{e}&=&\frac{\kappa}{c_q}=\frac{2 (d-2) (d-1) (z-2)\rh^{z-2}}{d^2}+\cdots.
\eea
Notice that as $d=2$ or $d=1$, the energy diffusion constant $D_{e}$ at leading order is vanishing and in this case, one shall consider the $\mathcal{O}(1/\beta^2_0)$ order instead.
The butterfly velocity multiplied by $\tau$ is given by
\be
v^2_{B}\tau=\frac{2\pi \rh^{z-2}}{(2-z)}.
\ee
Note that $\rh$ is a function $\beta_0$, $T$ and charge density. The ratios  $D_c/v^2_{B}\tau$ and $D_e/v^2_{B}\tau$ are all finite in this case.\\
$\bullet$ {Case II: $d-\theta+z-2\neq 0$, $(d^2-d\theta-dz+2\theta)= 0$}\\
The diffusion constants read
\bea\label{de}
D_{c}&=&\frac{\sigma}{\chi}=\frac{(d-2)\rh^{z-2}}{2 (d-1) (z-2)}+\cdots,\\
D_{e}&=&\frac{\kappa}{c_q}=\rh^{z-2}+\mathcal{O}(\beta^2_0)+\cdots.
\eea
The butterfly velocity multiplied by $\tau$ is given by
\be
v^2_{B}\tau=\frac{2 \pi  (d-2)\rh^{z-2}}{d (z-2)}.
\ee
The ration between the diffusion constants and $v^2_{B}\tau$ is given by $D_{c}/v^2_{B}\tau=\frac{d}{4 \pi  (d-1)}$ and $D_{e}/v^2_{B}\tau=\frac{d (z-2)}{2 \pi  (d-2)}$.\\
$\bullet$ {Case III: $d=\theta$, $z \neq 2$}\\
The diffusion constants in this case yield
\bea\label{de}
D_{c}&=&\frac{\sigma}{\chi}=\frac{\rh^{z-2}}{z-2}+\cdots,\\
D_{e}&=&\frac{\kappa}{c_q}=\frac{\rh^{z-2}(d-1) [( z-2 )\ln \rh-2] }{(d-\theta )}+\cdots.
\eea
The butterfly velocity multiplied by $\tau$ is given by
\be
v^2_{B}\tau=\frac{2 \pi  \rh^{z-2}}{(d-\theta)}.
\ee
 As $d\rightarrow \theta$ both the energy diffusion and the butterfly velocity diverges, but their ratio is finite $\frac{D_{e}}{v^2_{B}\tau}=\frac{(d-1) [( z-2 )\ln \rh-2] }{2 \pi } $. \\
$\bullet$ {Case IV: $d=1$, $z = 1+\theta$}\\
The diffusion constants are given by
\bea\label{de}
D_{c}&=&\frac{\sigma}{\chi}=-\rh^{z-2}\ln\rh+\cdots,\\
D_{e}&=&\frac{\kappa}{c_q}=\frac{\rh^{z-2} }{2(2-z)}+\cdots.
\eea
One can easily verify that as $d=1$, $z=1$ and $\theta=0$ and in the strong incoherent limit, the diffusion constants $D_c$ and $D_e$ recover the ansatz given in the previous section (\ref{66}-\ref{67}). This reflects that the calculations in this section is consistent with the previous section.
The butterfly velocity multiplied by $\tau$ is given by
\be
v^2_{B}\tau=\frac{2 \pi  \rh^{z-2}}{(2-z)}\rightarrow {\rm finite}.
\ee
We can see that the ratio between $D_{e}$ and $v^2_{B}\tau$ is finite $D_{e}/v^2_{B}\tau= \frac{1}{4\pi}$. \\
$\bullet$ {Case V: $d=\theta$, $z = 2$}\\
The diffusion constants are obtained as
\bea\label{de}
D_{c}&=&\frac{\sigma}{\chi}=-\ln\rh+\cdots,
\eea
We can express the charge diffusion in terms of the black hole temperature $D_{c}=\ln 2-2\pi T/\beta_0$, where we have used $\rh=\pi T+\sqrt{q^2_2+4\pi^2 T^2+\beta^2_0}/2$.
In this case the specific heat $c_q$ is zero and the energy diffusion might be ill-defined.
The butterfly velocity times $\tau$
\be
v^2_{B}\tau=\frac{2 \pi  \rh^{z-2}}{(d-\theta)}\rightarrow \infty.
\ee
Notice that in this case, the metric takes a very special form
\bea
ds^2&=&-r^2 f(r)dt^2+\frac{dr^2}{r^4 f(r)}+dx^2,\\
f(r)&=& 1-\frac{\rh^2}{r^2}+\frac{(q^2_2+\beta^2_0)}{r^2}\ln \frac{\rh}{r},\\
F_{(1)rt}&=&q_1 r, ~~~F_{(2)rt}=q_2 r^{-1}.
\eea
The entropy density in this case is $s=1/4G$ and the Hawking temperature is $T=\frac{\rh}{2\pi}(1-\frac{q^2+\beta^2_0}{2\rh^2})$.  The specific heat  becomes vanishing since the entropy density takes a constant value.
The universal dc electric conductivity $\prod_{i}\sigma_{ii}|_{q_i=0}=\prod_{i}Z^{d}_i \mathcal{A}^{d-2}$ is satisfied for all the above five cases.
\\
 $\bullet$ {Diffusivity in the absence of disorder parameter}\\
This is a very intriguing situation that even without translational symmetry breaking (i. e. $\beta=0$), finite DC electric conductivity can still be realized because of the presence
of the auxiliary $U(1)$ charge $q_1$ [64]. In the strong auxiliary gauge field  limit $q_1\rightarrow \infty$, the charge diffusion constant $D_c$ reads
\be
D_c=\frac{\rh^{z-2}}{d-\theta+z-2}.
\ee
The energy diffusion constant $D_e$ is given by
\be
D_e=\frac{(d+z-\theta)z}{(d-\theta)q^2_1}\rh^{z+\theta-d-2\theta/d}+\frac{q(d+z-\theta-2)(2d+z-2-2\theta)}{q^2_1(d-\theta)}\rh^{2-3d-z+3\theta-2\theta/d}.
\ee
In this case, the behavior of the energy diffusion is different from the charge diffusion. Notice that the non-vanishing of the auxiliary charge density $q_1$ is a consequence of $z\neq 1$.

We then consider a special $2+1$-dimensional black hole solution with a non-zero hyperscaling violating factor. Note that as $d=1$ and $d+z-\theta-2= 0$, we have a very particular
black hole solution as
\bea \label{hyperbtz}
ds^2&=&r^{-2\theta}\bigg(-r^{2z}f(r)dt^2+\frac{dr^2}{r^2f(r)}+r^2 dx^2\bigg),\\
f(r)&=&1-\frac{M}{r^{1+z-\theta}}-\frac{q^2_2+\beta^2_0}{2(1-\theta)r^{1+z-\theta}}\ln r,\nonumber\\
&=& 1-\frac{\rh^2}{r^2}-\frac{q^2_2+\beta^2_0}{2(1-\theta)r^2}\ln \frac{\rh}{r},\\
F_{(1)rt}&=&q_1 r, ~~~F_{(2)rt}=q_2 r^{-1},~~~ Z_1=r^{-2}, ~~~Z_2=1.
\eea
% The scalar potentials are given by \be A_{(1)t}=\mu_1\bigg(1-\frac{r^2}{\rh^2}\bigg),~~~A_{(2)t}=q_2 \ln \frac{r}{\rh}
%\ee
%The chemical potentials and charge density are given by $\mu_1$ and $\mu_2=q_2\ln(\rh)$.
The event horizon locates at $r=\rh$ satisfying $f(\rh)=0$. The Hawking temperature is given by
\be
T=\frac{\rh^{1+\theta}}{2\pi}\bigg(1-\frac{q^2_2+\beta^2_0}{4(1-\theta)\rh^2}\bigg).
\ee
As $\theta=0$ and then $q_1=0$, we recover the standard BTZ black hole solution obtained in (\ref{btz}) since $\theta=0$ and $d=1$ infers $z=1$.
The entropy density and the butterfly velocity are
\be
s=\frac{\rh^{1-\theta}}{4G}, ~~~v^2_B=\frac{2\pi T}{(1-\theta)\rh^{1-2\theta}}.
\ee
The transport coefficients are obtained as
\bea\label{sigmamatrix}
\sigma_{11}&=&\frac{1}{\rh^{3-\theta}}+\frac{q^2_1}{\beta^2_0 \rh^{1-\theta}},~~~ \sigma_{12}=\sigma_{21}=\frac{q_1 q_2}{\beta^2_0 \rh^{1-\theta}},\\
\sigma_{22}&=&\rh^{\theta-1}+\frac{q^2_2}{\beta^2 \rh^{1-\theta} }, ~~~\alpha_1=\frac{4\pi q_1}{\beta^2_0}, \alpha_1=\frac{4\pi q_2}{\beta^2_0},\\
\bar{\kappa}&=&\frac{16 \pi^2 T \rh^{1-\theta} }{\beta^2_0}.
\eea
We observe that as $\theta \rightarrow 0$, the quantity $q_1$ becomes vanished but the DC conductivity $\sigma_{11}=\rh^{-3}$ is not vanishing. The situation is very subtle in that
if we set $z=1$ and $\theta=0$ in the action (\ref{action1}), then the auxiliary gauge field,  $\sigma_{11}$ and $\sigma_{12}$ do not appear at the all.  This reflects that once the fluctuations of the auxiliary gauge field is turned on, there exists a discontinuity in the $\theta \rightarrow 0$, $(d+z-\theta)\rightarrow 2$ and $q_1 \rightarrow 0$ limit since $\sigma_{11}$  does not vanish in this limit \cite{lv16}.

On the other hand, as $\theta=-1$, the black hole solution shows its strange behaviors since $\rh^2<(q^2+\beta^2_0)/8$ corresponds to $T<0$ and $\rh^2\geq (q^2+\beta^2_0)/8$ corresponds to $T\geq 0$. Moreover, the black hole has its maximal Hawking temperature $T_{max}=\frac{1}{2\pi}$  as $\rh^2 \gg (q^2+\beta^2_0)/8$. Moreover, the null energy condition is violated as $\theta=-1$ and $d=1$. It seems that this black hole solution is not very physical.

%In what follows, we will not discuss the $\theta=-1$ case.

%The only non-trivial case for non-zero $\theta$ under the condition $d=1$ is $\theta=d=1$ and $z=2$, which the null energy condition is saturated.  The metric takes the form
%\bea
%ds^2&=&-r^2 f(r)dt^2+\frac{dr^2}{r^4 f(r)}+dx,\\
%f(r)&=& 1-\frac{\rh^2}{r^2}+\frac{(q^2_2+\beta^2_0)}{r^2}\ln \frac{\rh}{r},\\
%F_{(1)rt}&=&q_1 r, ~~~F_{(2)rt}=q_2 r^{-1}.
%\eea
%The entropy density in this case is $s=1/4G$ and the Hawking temperature is $T=\frac{\rh}{2\pi}(1-\frac{q^2+\beta^2_0}{2\rh^2})$. The butterfly velocity $v^2_B=\frac{2\pi T \rh^{z-2}}{(d-\theta)}$ becomes divergent as $d\rightarrow \theta$. On the other hand, the specific heat  also diverges since the entropy density takes a constant value. So, does the relation $D\sim v^2_B \tau$ still holds for this special black hole solution ?

 %Interestingly, as $d \rightarrow 1$, $z\rightarrow 2$ and $\theta\rightarrow 1+\varepsilon$, the charge diffusion constant behave as
%\be
%D_{c}=\frac{v^2_B \tau}{2\pi}
%\ee
%But the energy diffusion is vanishing as $D_{e}\sim {\varepsilon}$ in the $\varepsilon\rightarrow 0$ limit.

\section{Conclusion and discussions}

In summary, we obtain a class of black hole solutions analogous to charged BTZ black holes by considering $d+2$-dimensional action with non-trivial Lifshitz dynamical exponent $z$ and hyperscaling violating factor $\theta$. Those BTZ-like black hole solutions can be realized  because special combinations of $d$, $z$ and $\theta$ lead to divergence of the mass-, charges- and axions-related terms. Such divergences can be annihilated by renormalizing  the mass parameter. As summarized in table \ref{table},  there are five concrete cases that such charged BTZ-like black hole solutions can be realized.

We then show that the action of the charged BTZ black hole can be reduced to the Jackiw-Teitelboim theory by dimensional reduction. We find that the relation $D\sim v^2_B \tau$ is well obeyed by the standard charged BTZ black holes in the incoherent limit. We thus study the diffusions for general $d$, $z$ and $\theta$ and obtain general expressions for the charge and the energy diffusions. We carefully evaluate the diffusions for those five special cases. We can see that for cases I and II, the ratio $D_c/v^2_B\tau$ is finite, while $D_e \sim v^2_B\tau$ is valid for cases I, II, III, IV and V. In this sense, the energy diffusion seems more general than the charge diffusion. However, for case V, the charge diffusion is finite and the energy diffusion seems ill-defined, while $v^2_B\tau$ is divergent. Since there are two $U(1)$ gauge fields in the theory, we calso calculate the diffusion constants in the absence of momentum relaxation parameter $\beta$. In this case, the charge diffusion is same as that of the momentum dissipated case, but the energy diffusion has $\mathcal{O}(1/q^2_1)$ dependence.

We also examine the universial electrical DC conductivity formula and find that for Lifshitz spacetime with auxiliary $U(1)$ gauge fields, this formula is satisfied.
\section*{Acknowledgement} The study was partially supported by NSFC,
China (grant No.11375110); NSFC (grant No. 11475179); the Ministry of Science and Technology (grant No. MOST 104-2811-M-009-068) and National Center for Theoretical Sciences in Taiwan; and by NSFC
China (grant No.11275120). SJS was supported by the NRF, Korea (NRF-2013R1A2A2A05004846).\\

\end{document}